\documentclass[prl,twocolumn,showpacs,amsmath,amssymb]{revtex4}

\usepackage{graphicx}% Include figure files
\usepackage{dcolumn}% Align table columns on decimal point
\usepackage{bm}% bold math

\begin{document}

\title{Characteristic electronic structure and its doping evolution in lightly-doped to underdoped YBa$_2$Cu$_3$O$_y$}

\author{H. Yagi$^1$, T. Yoshida$^1$, A. Fujimori$^1$, K. Tanaka$^2$, N. Mannella$^2$, W. L. Yang$^2$, X. J. Zhou$^2$, D. H. Lu$^2$, Z.-X. Shen$^2$, Z. Hussain$^3$, M. Kubota$^4$, K. Ono$^4$, K. Segawa$^5$, Y. Ando$^5$, D. Iijima$^1$, M. Goto$^1$, K. M. Kojima$^1$, and S. Uchida$^1$}
 \affiliation{$^1$Department of Physics, Univerity of Tokyo, Bunkyo-ku, Tokyo 113-0033, Japan
\\$^2$Department of Appplied Physics and Stanford Synchrotron Radiation Laboratory, Stanford University, Stanford, CA 94305
\\$^3$Advanced Light Source, Lawrence Berkeley National Lab, Berkeley, CA 94720
\\$^4$Institute of Materials Structure Science, Tsukuba, Ibaraki 305-0801, Japan
\\$^5$Institute of Scientific and Industrial Research, Osaka University, Mihogaoka 8-1, Ibaraki, Osaka 567-0047, Japan
}

\date{\today}

\begin{abstract}
We have performed an angle resolved photoemission spectroscopy (ARPES) study of lightly-doped to underdoped YBa$_2$Cu$_3$O$_y$ (YBCO) untwinned single crystals and a core-level x-ray photoemission spectroscopy (XPS) study of YBCO single and polycrystals.
In the zone diagonal (nodal) direction, dispersive quasi-particle (QP) features crossing the Fermi level were observed down to the hole concentration of $\sim 4\% $, which explains the metallic transport of the lightly-doped YBCO.
The chemical potential shift estimated from XPS was more rapid than in the Bi2212 cuprates.
Upon hole doping, very rapid spectral weight transfer from high binding energies to the QP feature, even faster than La$_{2-x}$Sr$_{x}$CuO$_{4}$, was observed.
\end{abstract}


\maketitle

\section{Introduction}
The doping evolution of the electronic properties of lightly-doped cuprates have attracted considerable interest because of their fundamental importance in the context of both physics near Mott transition and mechanism of superconductivity..
In the cuprates, in particular, the phase competition and the nature of the competing phases in the underdoped region have provided major experimental and theoretical challenges.
Angle-resolved photoemission spectroscopy (ARPES) studies of lightly-doped La$_{2-x}$Sr$_{x}$CuO$_{4}$ (LSCO)~\cite{Yoshida_LSCO} have shown that a sharp dispersive quasi-particle (QP) peak crosses the Fermi level ($E_{\rm F}$) in the nodal $\vec{k}$ = (0,0)-($\pi$,$\pi$) direction, which explains their metallic behavior~\cite{LSCO_metallic} and that the other part of the Fermi surface around ($\pi$,0) is gapped or pseudo-gapped, resulting in an ``arc'' of the Fermi surface around the nodal direction.
A remnant of the lower Hubbard band (LHB) (or the polaronic side band~\cite{Roesch}) persists at $\sim -$(0.4-0.6) eV as the QP peak appears near $E_{\rm F}$.
As holes are further doped, spectral weight is transferred from the ``LHB'' to the QP peak~\cite{LSCO_mu}.
ARPES studies on lightly-doped Ca$_{2-x}$Na$_x$CuO$_2$Cl$_2$ (Na-CCOC)~\cite{kfCCOC} and Bi$_2$Sr$_{2-x}$La$_x$CuO$_{6+\delta}$ (Bi2201)~\cite{Bi2201_mu}, on the other hand, have shown quite different behaviors:
With hole doping, the ``LHB'' feature is shifted toward $E_{\rm F}$ and merges into the QP band crossing $E_{\rm F}$, making the LHB-to-QP spectral weight transfer less obvious.
These observations are consistent with the pinning of the chemical potential in LSCO and the rapid chemical potential shift in Na-CCOC, Bi2201 and Bi2212 as implied by core-level x-ray photoemission spectroscopy (XPS)~\cite{LSCO_mu,Bi2201_mu,Bi2212_mu,CCOC_mu}.

Recent in-plane transport studies of YBa$_2$Cu$_3$O$_y$ (YBCO) have revealed their metallic behavior ($d\rho/dT>0$) at high temperatures even in the lightly-doped ``insulating'' region as in the case of LSCO~\cite{Ando_transport}.
The $E_{\rm F}$ crossing of QP peak in the nodal direction has been observed in lightly-doped~\cite{YBCO_nodalQP} as well as in underdoped to overdoped YBCO~\cite{YBCO_bilayer}.
It is therefore important to see whether the electronic structure of lightly-doped YBCO is similar to that of LSCO or to that of Na-CCOC, Bi2201, and Bi2212.
ARPES studies of YBCO have not been advanced compared to those of other high-$T_c$ cuprates because intense surface-state signals near $E_{\rm F}$ mask the bulk electronic states around the X [=($\pi$,0)] and Y [=(0,$\pi$)] points in the Brillouin zone.
Lu \textit{et al.}~\cite{ARPES_Lu} have made high-resolution ARPES measurments on YBa$_2$Cu$_3$O$_{6.993}$ and have succeeded in distinguishing bulk electronic states from the surface states around X and Y by aging the sample surfaces and deduced the superconducting gap.
They have reported that the superconducting peak is accompanied by a peak-dip hump structure, similar to Bi2212.
Yoshida \textit{et al.}~\cite{YBCO_nodalQP} and Borisenko \textit{et al.}~\cite{YBCO_bilayer} have reported a bilayer splitting in lightly-doped and optimally-doped YBCO, respectively.
In a recent ARPES study of optimally-doped YBCO by Nakayama \textit{et al.}~\cite{YBCO_Takahashi}, bulk and surface Fermi surfaces have been separated. More recently, Hossain \textit{et al.} reported the doping evolution of the electronic structure of the topmost CuO$_2$ plane from overdoped to the underdoped region~\cite{YBCO_Hossain}.
The recent Hall resistance measurment of underdoped YBCO revealed the existence of an small electron pocket~\cite{QO1_YBCO,QO2_YBCO}, although in the ARPES studies of other high-$T_{\rm c}$ cuprates, only a Fermi arc around the nodal direction has been observed.
Therfore, it is important to clarify whether an small electron pocket is observed in ARPES studies of underdoped YBCO or not.

In this paper, ARPES and core-level XPS studies of lightly-doped to underdoped YBCO are presented to clarify the above issues and to gain further insight into the electronic structure of YBCO.

\begin{table}
\caption{Oxygen content $y$, hole concentration $\delta$ in the CuO$_2$ plane estimated from the electrical resistivity
and thermopower~\cite{Ong&Uchida}, and $T_c$ of the YBCO samples studied in the present work.}
\label{tab:table1}

\begin{tabular}{p{2.5cm}|p{2.5cm}|p{2.5cm}}
\hline
\hline
$y$ & $\delta$ & $T_c$ (K) \\
\hline
6.20 & 0.01 &  \\
6.25 & 0.02 &  \\
6.28 & 0.03 &  \\
6.30 & 0.04 &  \\
6.35 & 0.05 &  \\
6.40 & 0.07 &  \\
6.45 & 0.08 & 20 \\
6.50 & 0.09 & 35 \\
6.60 & 0.12 & 57 \\
\hline
\hline
\end{tabular}

\end{table}
\section{Experiment}
Untwinned single crystals of YBa$_2$Cu$_3$O$_y$ ($y$=6.25, 6.28, 6.30, 6.35, 6.40, 6.45, 6.60) were grown by the flux method as described in Ref.~\cite{sample_growth}.
In the XPS study, we measured polycrystals of YBa$_{2}$Cu$_{3}$O$_y$
($y=$6.20, 6.40, 6.50, 6.60), too.
In YBCO, how the in-plane hole concentration $\delta$ changes with oxygen content $y$ has been controversial.
We have adopted the estimates from electrical resistivity and thermopower~\cite{Ong&Uchida}.
$y$, $\delta$ and $T_c$ of the measured samples are listed in Table~\ref{tab:table1}.
The ARPES measurements were carried out at BL10.0.1 of Advanced Light Source (ALS) using incident photons of 55 eV and a SCIENTA R4000 analyzer, at BL28 of Photon Factory (PF) using incident photons of 65 eV, a SCIENTA SES2002 analyzer, and at BL5-4 of Stanford Synchrotron Radiation Laboratory (SSRL) using photons of 28 eV and a SCIENTA SES-200 analyzer.
An R-Dec Co. Ltd. i GONIO LT goniometer~\cite{igonio} was used at PF.
The total energy and momentum resolution was about 20 meV and 0.02$\pi$ in units of 1/\textit{a} ($a=3.86~$\AA~is the lattice constant), respectively.
The samples were cleaved \textit{in situ} in an ultra high vaccum of $10^{-11}$~Torr and cooled to $\sim$ 10 K during the measurement.
The XPS measurements were carried out using the Mg $K{\alpha}$ line (${\it h}{\nu}=1253.6$~eV) and a VSW EAC 125 hemispherical analyzer for sintered polycrystalline samples.
Single cyrstals were measured using the Mg $K{\alpha}$ line (${\it h}{\nu}=1253.6$~eV) and a SCIENTA SES-100 analyzer.
The energy resolution was about $0.8$~eV, which was largely due to the width of the photon source.
Nevertheless we could determine the core-level shifts with an accuracy of about $\pm$50 meV because most of the spectral line shapes were identical between different compositions.
Details of the core-level XPS studies are described elsewhere~\cite{mu_review,CCOC_mu}.

\section{Results and Discussion}
\begin{figure}[htbp]
\begin{center}
\includegraphics[width=9.5cm]{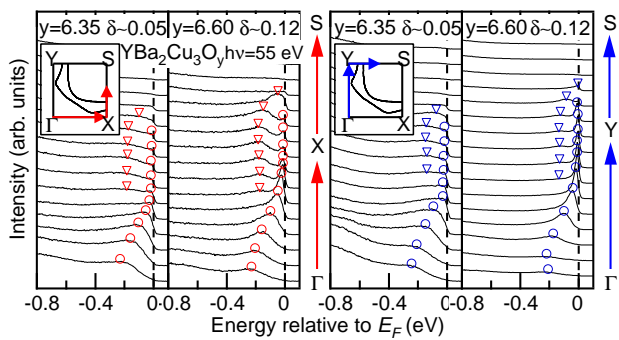}
\caption{(Color online) EDC's of lightly-doped to underdoped YBCO. Two features in the spectra are surface bonding state (open triangles) and surface antibonding state (open circles).}
\label{bilayer_EDC}
\end{center}
\end{figure}

Figure \ref{bilayer_EDC} shows energy distribution curves (EDC's) along high symmetry lines in the second Brillouin zone for the $y$=6.35 and 6.60 samples.
Two dispersive features are observed and both of them become more pronounced in going from y=6.35 to 6.60, reflecting the increased hole doping.
One (open circles) is around $-0.2$ eV at $\Gamma$, dispersing upward and crossing $E_{\rm F}$ around 60 \% of the $\Gamma$X and $\Gamma$Y lines.
In optimally-doped YBCO, this feature was assigned to surface states in Ref.~\cite{ARPES_Lu} and assigned to surface antibonding band in Ref.~\cite{YBCO_Takahashi}
The other feature (open triangles) was considered to be the surface bonding band according to Ref.~\cite{YBCO_Takahashi}.
The dispersion of the surface bonding band is unclear around $\Gamma$ because the broad and strong antibonding band masks this feature.
It should be noted that the energy position of the surface bonding band showed clear anisotropy, $-0.18$ eV at X and $-0.14$ eV at Y.
The 1D like chain-derived states which was observed in the spectra of optimally doped YBCO~\cite{ARPES_Lu} is not observed here, perhaps due to the oxygen deficiency in the CuO chain and also due to the difference in the photon energy and the polarization of the incident light, i.e., difference in the matrix elements.

\begin{figure}[htbp]
\begin{center}
\includegraphics[width=8cm]{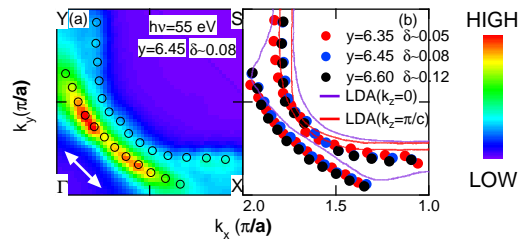}
\caption{(Color online) (a) Spectral weight mapping at $E_{\rm F}$ for a 40 meV integration window. Open circles are the $k_{\rm F}$ positions determined from MDC's and the white arrow denotes the polarization of the incident light. (b) Experimentally determined $k_{\rm F}$ positions in each doping level and LDA Fermi surface for $k_{\rm z}$=0 and $k_{\rm z}$=$\pi$/$c$ related to the CuO$_2$ plane.}
\label{bilayer_mapping}
\end{center}
\end{figure}

Figure \ref{bilayer_mapping}(a) shows the plot of spectral weight in the second Brillouin zone integrated within 40 meV of $E_{\rm F}$, which reflects the Fermi surface, for $y$=6.45.
Not only a large hole-like Fermi surface centered at the S point, but also an apparently electron-like Fermi surface centered at $\Gamma$ are observed, indicating the bilayer splitting~\cite{YBCO_nodalQP,YBCO_bilayer}.
Open circles in Fig.~\ref{bilayer_mapping}(a) are the $k_{\rm F}$ positions
of the Fermi surfaces determined by the peak positions of momentum distribution curves (MDC's) at $E_{\rm F}$.
In Fig.~\ref{bilayer_mapping}(b), the $k_{\rm F}$ positions thus determined are plotted for various compositions in the same panel.
The shape of the Fermi surface does not vary appreciably in this doping range, suggesting that these Fermi surfaces are not derived from bulk.

Because the cleaved surface of YBCO tend to be overdoped~\cite{YBCO_bilayer,YBCO_Takahashi}, we compare the experiment with the Fermi surfaces predicted by local-density-approximation (LDA) calculation for $y$=7~\cite{LDA_Andersen} in Fig.~\ref{bilayer_mapping}(b).
In the LDA calculation, the Fermi surface has four sheets derived from the CuO$_2$ planes and the CuO chain.
However, because we did not observe the CuO chain-derived bands, we have plotted only the CuO$_2$ plane-derived sheets, that is, those derived from the bonding and antibonding bands of the CuO$_2$ bilayer.
The Fermi surfaces for two values of $k_{\rm z}$=0 and $\pi/c$ are indicated because for the optimally-doped YBCO the CuO chain mediates orbital overlap along the $c$-axis and therefore strong $k_{\rm z}$ dispersions are predicted.
The experimental results are more similar to the $k_{\rm z}$=0 cross-section than to the $k_{\rm z}$=$\pi/c$ one.
The shape of the Fermi surface of the bonding band is more similar to that of Bi2212 than to that of LSCO.
According to the tight binding model including hopping parameters $t$, $t'$, and $t''$, this suggests a large next-nearest-neighbour hopping $|t'|$ in YBCO~\cite{Tohyama_SST}.

\begin{figure}[htbp]
\begin{center}
\includegraphics[width=8cm]{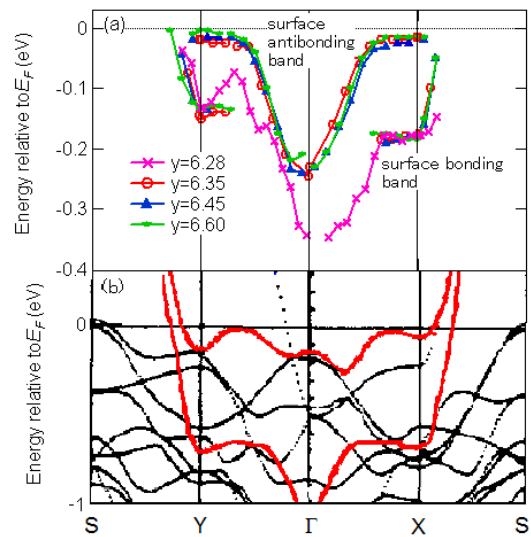}
\caption{(Color online) Band dispersions in YBCO obtained from EDC's (a) compared with the LDA band dispersions for YBCO (y=7) at $k_{\rm z}$=0~\cite{LDA_Andersen} (b). In (b), red curves are the antibonding and bonding bands of the CuO$_2$ planes.}
\label{bilayer_band}
\end{center}
\end{figure}

In Fig.~\ref{bilayer_band}(a), we summarize the surface band dispersions of YBCO for various $y$'s obtained from EDC's.
Here, the surface bonding band for the $y$=6.28 sample was taken in a different experimental geometry ($\bm{E}\perp \bm{b}$) and photon energy (${\it h}{\nu}=65$~eV) than the rest of the data ($\bm{E}\parallel \rm{XY}$), and was clearly resolved even around the $\Gamma$ point.
Although the spectral line shape and intensities dramatically changed with hole doping (see Fig.~\ref{bilayer_EDC}), the band dispersions did not change appreciably in this doping range.
The lower bound for the bilayer splitting at X and Y is estimated to be 180 meV and 140 meV, respectively, because the surface antibonding band should be above $E_{\rm F}$ at these momenta.
Figure~\ref{bilayer_band}(b) shows the LDA band dispersions of YBCO ($y$=7) for $k_{\rm z}$=0~\cite{LDA_Andersen}.
Red curves are the bonding and antibonding bands of the CuO$_2$ bilayer.
The experimental surface band dispersions are qualitatively reproduced by the LDA band structure except for the $E_{\rm F}$ crossing points along the symmetry lines and the overall band narrowing (by a factor of $\sim$ 0.5) in experiment.
This similarity indicates that the origin of the surface state is the overdoped  cleaved surface.~\cite{YBCO_Takahashi}
In the LDA calculation, the bonding-antibonding splitting of the CuO$_2$ bilayers at X and Y was 610 meV and 560 meV, respectively, which is much larger than the experimental value of $\gtrsim$ 180 meV and $\gtrsim$ 140 meV.
The reason for this dicrepancy is that LDA calculation neglects electron correlation and generally overestimates band dispersions.
Indeed, overdoped Bi2212 has shown a bilayer splitting of 88 meV~\cite{Feng}, much smaller than 300 meV predicted by LDA calculation~\cite{Chakravarty,LDA_Andersen}.
Bilayer Hubbard model calculations for two coupled CuO$_2$ layers, which explicitly include the on-site Coulomb repulsion, predicted a maximum splitting of 40 meV~\cite{bilayer_Hubbard}.

Figure 4(a)-(c) shows spectral weight mapping at $E_{\rm F}$ with a 40 meV integration window in the first Brillouin zone for the $y$=6.35, 6.45 and 6.60 samples.
A Fermi "arc" around the nodal direction is observed as in the LSCO case, which is completely different from Fig.~2(a).
Figure 2(a) does not show a Fermi arc but a hole-like and an electron-like Fermi surface arising from the surface state.
Figure 4(e) shows the intensity plot in the \textit{E}-$\bm{k}$ plane in the nodal direction for the $y$=6.35 sample together with the energy band dispersion determined from MDC peaks.
The dispersion of this band shows a kink structure around $\sim$ 50 meV like in many other high-$T_c$ cuprates, while the surface bonding band is reported to show a straight dispersion~\cite{YBCO_Takahashi}.
In addition, the $k_{\rm F}$ positions in the nodal direction in the first Brillouin zone are different from those in the second Brillouin zone.
These findings suggest that the Fermi arc observed in the first Brillouin zone is distinct from the Fermi surfaces observed in the second BZ and we attribute it to a bulk-derived feature.
To investigate the doping evolution of the Fermi arc, the $k_{\rm F}$ positions determiend by the peak positions of the MDC's are plotted by filled circles in Fig.~4(a)-(c).
The $k_{\rm F}$ positions for $y$=6.35, 6.45, and 6.60 are overlaid in Fig.~4(d).
As holes are doped, the hole-like Fermi surface centered at S slightly expands in the nodal region, which is also consistent with the assumption that these Fermi surfaces are bulk derived.
\begin{figure}[htbp]
\begin{center}
\includegraphics[width=8cm]{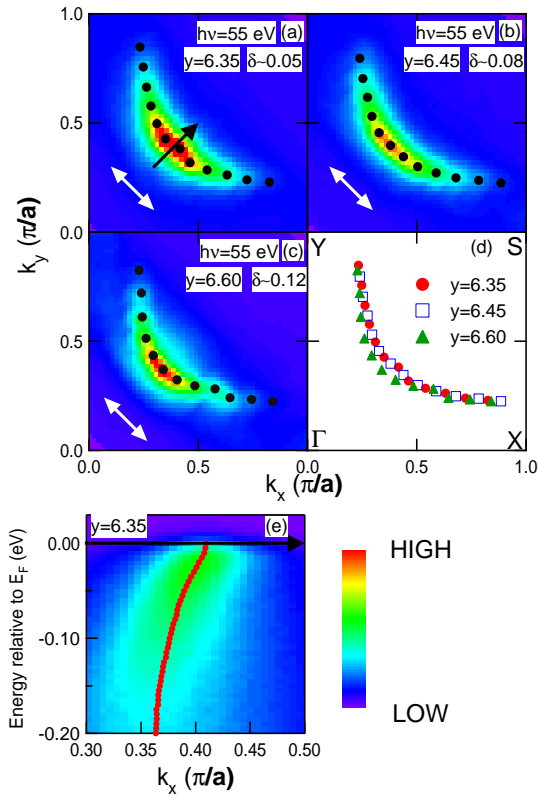}
\caption{(Color online) (a)-(c) Spectral weight mapping at $E_{\rm F}$ for a 40 meV integration window in the first Brillouin zone. Filled circles are the $k_{\rm F}$ positions determined from MDC's and the white arrow denotes the polarization of the incident light. (d) $k_{\rm F}$ positions for $y$=6.35, 6.45, and 6.60 determined from MDC's. (e) Intensity plot in the \textit{E}-$\bm{k}$ plane in the nodal direction for the $y$=6.35 sample. Filled circles denote the energy band dispersion determined by fitting the MDC's to Lorentzians.}
\label{Fig4}
\end{center}
\end{figure}

\begin{figure}[htbp]
\begin{center}
\includegraphics[width=6cm]{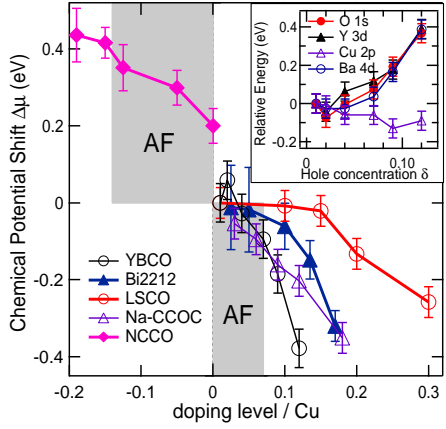}
\caption{(Color online) Chemical potential shift in YBCO compared with those in other high $T_c$ superconductors~\cite{LSCO_mu,CCOC_mu,Bi2212_mu,NCCO_mu}.
The antiferromagnetic region is displayed by shaded region for YBCO and NCCO. Inset shows the shift of each core level for YBCO as a function of hole concentration $\delta$.}
\label{chemicalpotential}
\end{center}
\end{figure}

Next, we present the result of the measurements of XPS core levels in YBCO, and deduce the chemical potential shift as a function of doping.
As shown in the inset of Fig.~\ref{chemicalpotential}, the O 1\textit{s} and Y 3\textit{d} core levels show the same shifts.
As for the Ba 4\textit{d} core level, the peak becomes broader with hole doping, perhaps due to some surface effects, but apart from this the Ba 4\textit{d} peak position follow the shifts of the O 1\textit{s} and Y 3\textit{d} core levels, too.
The line shape of the Cu 2\textit{p} peak was not identical between the different compositions but its peak position moved in the opposite direction to those of the O 1\textit{s}, Y 3\textit{d} and Ba 4\textit{d} core levels.
The different behavior of the Cu 2\textit{p} core level can be attributed to the creation and the growth of ``Cu$^{3+}$" signals with hole doping on the higher binding energy side of the Cu$^{2+}$ main component~\cite{LSCO_mu}.
The inset of Fig.~\ref{chemicalpotential} shows the binding energy shifts referenced to $y$=6.2 of each core level thus estimated.
As in the previous studies~\cite{mu_review}, we consider that the parallel shifts of the O 1\textit{s}, Y 3\textit{d} and Ba 4\textit{d} core levels reflect the chemical potential shift $\Delta \mu$.
Recently, Maiti {\it et al.} reported the shifts of the core levels in YBCO using hard x-ray photoemission spectroscopy~\cite{Maiti}.
Our Y 3\textit{d} and O 1\textit{s} shifts are consistent with their results except for their lowest doped sample ($y$ = 6.15).
In the main panel of Fig.~\ref{chemicalpotential}, therefore, $\Delta \mu$ has been obtained by taking the average of the O 1\textit{s} and Y 3\textit{d} core-level shifts as shown in the figure.
In the same panel, the chemical potential shift in other high $T_c$ cuprates are plotted for comparison~\cite{LSCO_mu,CCOC_mu,Bi2212_mu,NCCO_mu}.
Theoretically, if the long-range antiferromagnetic order of the parent Mott insulator disappears for a very small hole concentration, the charge suceptibility $\partial n$/$\partial \mu$ diverges in the limit $\delta \to 0$~\cite{suppression} and hence the chemical potential is suppressed.
In LSCO and Bi2212, the antiferromagnetic order quickly disappears for a small amount of hole doping and indeed the shift is suppressed for small $\delta$.
In YBCO and NCCO, on the other hand, the antiferromagnetic order perisits up to $\delta \sim 0.07$ and $\delta \sim 0.14$, respectivly, and thefore it is reasonable that little or no suppression of the shift was observed for small $\delta$.

According to $t$-$t'$-$t''$-$J$ model calculations~\cite{cps_tj}, the larger the  $|t'|$ becomes, the stronger the chemical potential shift becomes.
The observed differences between Bi2212 and LSCO, such as the different dispersion width of the LHB from ${\bm k}\sim (\pi/2,\pi/2)$ to $\sim (\pi,0)$, the different shape of the Fermi surface as well as the different behavior of the chemical potential shift, can be explained by a larger $|t'|$ value in Bi2212 than that in LSCO~\cite{cps_tj,tj_tanaka}.
$\Delta \mu$ in YBCO is the largest among those compounds, which means that the $|t'|$ of YBCO may be the largest, consistent with the conclusion from the ARPES results described above.
The suppression of the shift in the underdoped LSCO is too strong compared to 
the $t$-$t'$-$t''$-$J$ model calculation and has been attributed to a fluctuation of charge stripes~\cite{LSCO_mu}.
The dynamical stripes seen in LSCO change their periodicity linearly with doping~\cite{stripe_LSCO2}.
As a result, the local charge density is nearly unchanged so that the hole rich region expands and the hole poor region shrinks as the average hole concentration is increased. 
Under such a circumstance, the chemical potential would not move with doping.
If the periodicity of the stripes remains unchanged with doping, on the other hand, the chemical potential is expected to move in a rigid-band-like manner.
In this regard, the observations of in-plane anisotropic magnetoresistance~\cite{Ando_magnetoresistance} and electrical resistivity~\cite{Ando_transport} in untwinned single crystals of YBCO are most consistent with the idea that charge stripes also exist in lightly-doped YBCO.
The presence of the stripes and the strong chemical potential shift can be reconciled if the stripes in the lightly-doped YBCO are of the type of fixed periodicity.

To investigate the doping dependence of the shape of the Fermi surface and the QP band quantitatively, we have fitted the $k_{\rm F}$ positions shown in Fig.~4(d) to the two-dimensional $t$-$t'$-$t''$ tight binding model
\begin{eqnarray*}
\epsilon _k - \mu = &-&2t[{\rm cos}(k_x a) + {\rm cos}(k_y a)]-4t'{\rm cos}(k_x a){\rm cos}(k_y a)\\ &-&2t''[{\rm cos}(2k_x a)+{\rm cos}(2k_y a)]+\epsilon _0
\end{eqnarray*}
Here, $\epsilon _0$ is the position of the band center relative to the chemical potential.
We have assumed that $t''/t' = -1/2$ as before~\cite{Yoshida_cond,cps_tj,Bi2201_mu}, and regarded $-t'/t$ and $- \epsilon _0 /t$ as fitting parameters.
Figure 6(a) shows the shape of the Fermi surface from the tight-binding fit.
One can see the squarelike Fermi surface which suggests large $-t'/t$~\cite{Tohyama_SST}.
Figure 6(b) shows fitted tight-binding parameters as function of hole concentration.
As in the case of LSCO and Bi2201, $-t'/t$ shows weak doping dependence~\cite{Yoshida_cond,Bi2201_mu}.
However, the $-t'/t$ value is much larger in YBCO ($\sim$ 0.4) than that in LSCO ($\sim$ 0.2) and Bi2201 ($\sim$ 0.2), consistent with the XPS results.
If the chemical potential moves in a rigid-band-like manner, $\Delta \mu/t = \epsilon _0 /t$, as is not the case in YBCO and LSCO.
The doping dependence of $\epsilon _0 /t$ in YBCO is similar to that in LSCO while $-\Delta \mu /t$ increases faster than that in LSCO.
This difference may be attributed to the different type of charge stripes as described above.

\begin{figure}[htbp]
\begin{center}
\includegraphics[width=8cm]{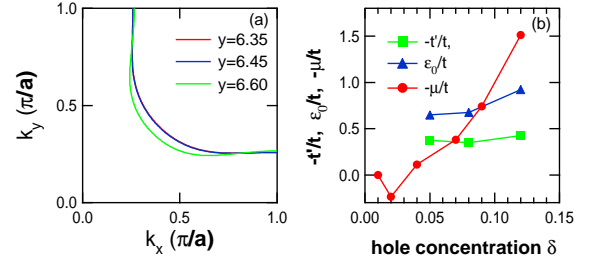}
\caption{(Color online) Doping dependence of the electronic structure in YBCO. (a) Doping dependence of the Fermi surface shape. (b) Doping dependence of the fitted tight-binding parameters.}
\label{tbparameters}
\end{center}
\end{figure}

\begin{figure}[htbp]
\begin{center}
\includegraphics[width=8cm]{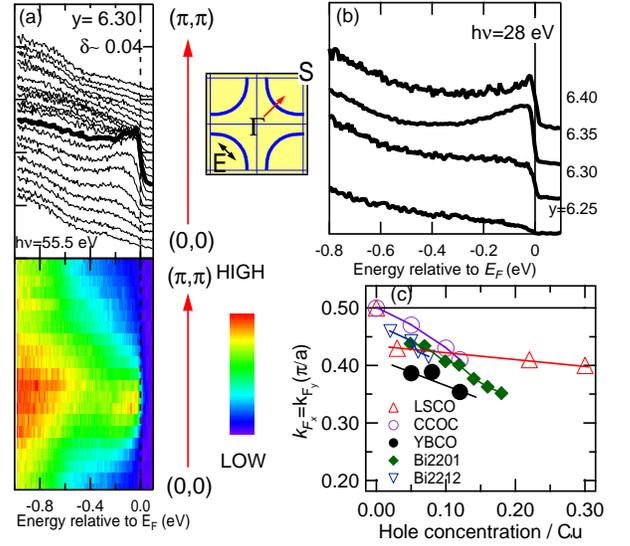}
\caption{(Color online) ARPES spectra taken at 10 K for lightly-doped
YBa$_2$Cu$_3$O$_y$ along the nodal direction in the first Brillouin zone.
$\delta$ indicates hole concentration per CuO$_2$ plane. (a) Top panel shows EDC's and bottom panel shows intensity plot in the \textit{E}-$\bm{k}$ plane. (b) EDC's at $k_{\rm F}$ for $y$=6.25, 6.30, 6.35, and 6.40 samples. (c) $k_{\rm F}$ positions for various doping levels in YBCO, LSCO~\cite{Yoshida_LSCO}, Na-CCOC~\cite{kfCCOC}, Bi2201~\cite{Bi2201_mu}, and Bi2212~\cite{Bi2212_kf}.}
\label{kf}
\end{center}
\end{figure}

Figure~\ref{kf}(a) shows EDC's (top) and corresponding intensity plots in \textit{E}-$\bm{k}$ space (middle) in the nodal (0,0)-($\pi$,$\pi$) direction of the $y$=6.30 sample.
As in the case of LSCO, a dispersive QP feature crossing $E_{\rm F}$ is seen for the lightly-doped $y\ge$6.30 ($\delta$=0.04) sample and its intensity increased with doping as shown in Fig.~\ref{kf}(b), giving rise to the metallic transport at high temperatures~\cite{Ando_transport}.
Figure \ref{kf}(c) shows the $k_{\rm F}$ positions along the nodal direction for various doping levels in YBCO, LSCO, Na-CCOC, Bi2201, and Bi2212.
In the antiferromagnetic (AF) insulating state of the undoped compound, the valence-band maximum along the nodal direction should be located at the AF zone boundary, that is, exactly at ($\pi/2,\pi/2$) due to the band folding in the AF state.
In Na-CCOC and the Bi-based compounds, indeed, with decreasing hole concentration the $k_{\rm F}$ position moves toward ($\pi/2,\pi/2$).
In LSCO, on the other hand, $k_{\rm F}$ extrapolates to $\sim$($0.44\pi ,0.44\pi$) and not to ($\pi/2,\pi/2$)~\cite{Yoshida_cond}.
In LSCO, in addition to the QP feature crossing $E_{\rm F}$, the broad ``LHB'' or the polaronic feature at $\sim$ -(0.4-0.6) eV is recognized for low hole concentrations $x\leq 0.05$.
Such an electronic structure has been regarded as the presence of the coherent and incoherent parts of the spectral function influenced by electron-electron and electron-phonon interacitons~\cite{Roesch} and/or due to a nano-scale electronic inhomogeneity such as charge stripes~\cite{Mayr}.
Since such an inhomogeneity is also expected to exist in lightly-doped YBCO~\cite{Ando_magnetoresistance,Ando_transport}, the fact that $k_{\rm F}$ in YBCO did not extrapolate to ($\pi/2,\pi/2$) suggests that QP band is also separated from the ``LHB'' in YBCO.
Unfortunately, such a remnant of the ``LHB'' is not seen in YBCO at least down to $\sim -$0.80 eV, partly because it is masked by the strong overcapping tail of the O 2\textit{p} band and partly because spectral weight transfer from the LHB to the QP band occurs very quickly with hole doping.

\section{Conclusion}
We have observed the band dispersions and Fermi surfaces of lightly-doped to underdoped YBCO by ARPES using untwinned single crystals.
The shape of the Fermi surface and the large chemical potential shift indicated large $|t'|$ like Bi2212, Bi2202, and Na-CCOC.
On the other hand, we observed a rather clear dispersive QP feature crossing $E_{\rm F}$ along the nodal direction, revealing a similarity to LSCO.
Although the coexistence of the remnant LHB could not be confirmed unlike the other cuprates, we conclude that very rapid spectral weight transfer ocurrs from the LHB to the QP band upon hole doping in the lightly-doped to underdoped YBCO.
We have also studied the chemical potential shift with hole doping by the XPS measurements of single and polycrystals.
The shift was as rapid as those in Na-CCOC, Bi2201, and Bi2212 in spite of the fact that charge stripes are observed in lightly-doped YBCO like LSCO.
We attribute the different chemical potential shifts between YBCO and LSCO to different types of charge stripes.

\section{Acknowledgement}
This work was supported by a Grant-in Aid for Scientific Research in Priority Area ``Invention of Anomalous Quantum Materials" (16076208) from MEXT, Japan.
ALS and SSRL is operated by the Department of Energy's Office of Basic Energy Science, Division of Chemical Sciences and Materials Sciences.
Experiment at Photon Factory was done under the approval of the Photon Factory Program Advisory Committee (Proposal Nos.~2006S2-001 and 2009S2-005).
The work at Osaka University was supported by KAKENHI Grant Nos.~19674002, 20030004, and 20740196.

\end{document}